\documentclass[conference]{IEEEtran}
\IEEEoverridecommandlockouts
\usepackage{cite}
\usepackage{amsmath,amssymb,amsfonts}
\usepackage{algorithmic,algorithm}
\usepackage{graphicx}
\usepackage{textcomp}
\usepackage{xcolor}
\usepackage{tabulary}
\usepackage{bm}
\def\BibTeX{{\rm B\kern-.05em{\sc i\kern-.025em b}\kern-.08em
    T\kern-.1667em\lower.7ex\hbox{E}\kern-.125emX}}

\newcommand{\bH}{\mathbf{H}}
\newcommand{\bh}{\mathbf{h}}
\newcommand{\bI}{\mathbf{I}}
\newcommand{\bw}{\mathbf{w}}
\newcommand{\bY}{\mathbf{Y}}
\newcommand{\by}{\mathbf{y}}

\newcommand{\bn}{\mathbf{n}}
\newcommand{\bN}{\mathbf{N}}

\newcommand{\bR}{\mathbf{R}}

\newcommand{\bvphi}{\boldsymbol{\varphi}}
\newcommand{\bphi}{\boldsymbol{\phi}}
\newcommand{\bPhi}{\boldsymbol{\Phi}}

\newcommand{\wh}{\widehat}
\newcommand{\ds}{\displaystyle}
\newcommand{\trace}{\text{tr}}
\newcommand{\APhi}{\mathbf{A}_{\boldsymbol{\Phi}}}
\newcommand{\norm}[1]{\left\lVert#1\right\rVert}

\renewcommand{\i}{\mathrm{i}}

\IEEEoverridecommandlockouts

\begin{document}

\title{RIS-aided Massive MIMO: Achieving Large Multiplexing Gains with non-Large Arrays}
\author{
\IEEEauthorblockN{Stefano Buzzi, Carmen D'Andrea and Giovanni Interdonato}
\IEEEauthorblockA{Dipartimento di Ingegneria Elettrica e dell'Informazione (DIEI),\\ University of Cassino and Southern Latium, Cassino, Italy \\
\{s.buzzi,carmen.dandrea, giovanni.interdonato\}@unicas.it}\thanks{The authors are also with the Consorzio Nazionale Interuniversitario
per le Telecomunicazioni (CNIT), 43124, Parma, Italy. This paper has been supported by the Italian Ministry of Education University and Research (MIUR) Project ``Dipartimenti di Eccellenza 2018-2022''. The work of C. D'Andrea has
been also partially supported by the ``Starting Grant 2020'' (PRASG) Research Project.}}

\maketitle

\begin{abstract}
This paper considers an antenna structure where a (non-large) array of radiating elements is placed at short distance in front of a reconfigurable intelligent surface (RIS). We propose a channel estimation procedure using different configurations of the RIS elements and derive a closed-form expression for an achievable downlink spectral efficiency by using the popular hardening lower-bound. Next, we formulate an optimization problem, with respect to the phase shifts of the RIS, aimed at minimizing the channels cross-correlations while preserving the channels individual norms. The numerical analysis shows that the proposed structure is capable of overcoming the performance of a conventional massive MIMO system without the RIS.
\end{abstract}


\section{Introduction}
Massive MIMO is one of the key technologies used to boost capacity in currently under deployment 5G wireless networks \cite{redbook}; indeed, leveraging the joint coherent transmission/reception from several antenna elements, and the fully digital beamforming at the base station (BS), the massive MIMO technology permits an aggressive spatial multiplexing of a large number of user equipments (UEs) in the same time-frequency slot. 

Recently, RISs have emerged as one of the most striking innovations for the evolution of 5G systems into 6G systems \cite{basar2019wireless,RuiZhang_TWC_2019}. A RIS is (usually) a passive planar structure, made of several sub-lambda spaced reflective elements, whose electromagnetic characteristics and in particular the phase offsets imposed when reflecting impinging waves can be controlled and adaptively changed. RISs can be thus employed to extend coverage, and/or to improve the signal-to-interference plus noise ratio (SINR) at the intended receiver location. Several studies have recently appeared showing the benefits that RISs can bring in different applications such as mobile edge computing networks\cite{Bai_MEC_JSAC2020}, physical layer security systems\cite{Hong_PLS_TCOM2020}, cognitive radio networks\cite{Zhang_Cognitive_TVT2020} and radar systems\cite{buzzi2021radar}.

In this paper, we consider the case in which a RIS is placed at short distance and in front of an active antenna array (see Fig. 1), with a non-large number of elements. The UEs to be served are placed in the backside of the active antenna array. The structure is to be intended as a low-complexity approximation of a massive MIMO system. A similar structure, but in a different setting and under different hypothesis, was also considered in \cite{jamali2020intelligent}. In the considered RIS-aided massive MIMO architecture, each active antenna is equipped with a dedicated RF chain and illuminates the RIS. Each passive element at the RIS receives a superposition of the signals transmitted by the active antennas and adds a desired phase shift to the overall signal. Leveraging the additional system optimization parameters introduced by the tunable reflecting elements of the RIS, a simpler transceiver architecture (i.e., less antennas and less RF chains) can be used than that usually employed in conventional massive MIMO systems with large-scale antenna array. 

We firstly provide a signal and channel model suitable for the analysis of such a system, then we propose a channel estimation procedure which, exploiting only the active antennas and using different configurations of the RIS phase shifts, is capable to estimate the channel from the large number of RIS elements to all the UEs.  We derive a closed-form expression for an achievable downlink spectral efficiency per user, by using the popular hardening lower-bound, and formulate a generic optimization problem which can be used for two purposes: i) the minimization of the cross-correlation among channel vectors to reduce the multiuser interference, and ii) striking a balance between the conflicting requirements that the RIS phases should not lower the norms of the composite channels, while they should minimize the cross-correlation among channel vectors. In the numerical results, we show the effectiveness of the considered architecture assuming both omnidirectional and directional active antennas which allow us to overcome the traditional system with only active antennas.
\section{System Model}

Consider a single-cell system where a base station is equipped with a RIS-aided antenna array, and serves $K$ single-antenna UEs. We denote by $N_A$ the elements of the active antenna array, and by $N_R> N_A$ the number of configurable reflective elements of the RIS. The uplink channel between the $k$-th UE and the RIS is represented by the $N_R$-dimensional vector ${\mathbf{h}}_k$, while the matrix-valued channel from the RIS to the receive antenna array is represented through the $(N_A \times N_R)$-dimensional matrix $\bH$. 
The tunable phase shifts introduced by the reflective elements of the RIS are represented through a diagonal $(N_R \times N_R)$-dimensional matrix $\bPhi$; in particular, the $\ell$-th coefficient on the diagonal of $\bPhi$ is $e^{\i \phi_\ell}$, with $\phi_\ell$ the phase shift associated to the $\ell$-th element of the RIS. 
Accordingly, the composite uplink channel from the $k$-th UE to the active array is the $N_A$-dimensional vector $\bar{\bh}_k= \bH \bPhi \bh_k$. 

Under the rich-scattering assumption, usually verified at sub-6 GHz frequencies, the channel vector $\bh_k$ is modeled as a complex Gaussian distributed vector with zero mean and covariance matrix $\beta_k \bI_{N_R}$, with $\bI_{N_R}$ denoting the identity matrix of order $N_R$ and $\beta_k$ representing the large-scale fading coefficient for the $k$-th UE. 

\subsection{Channel Model}
\begin{figure}[t]
	\begin{center}	\includegraphics[scale=0.3]{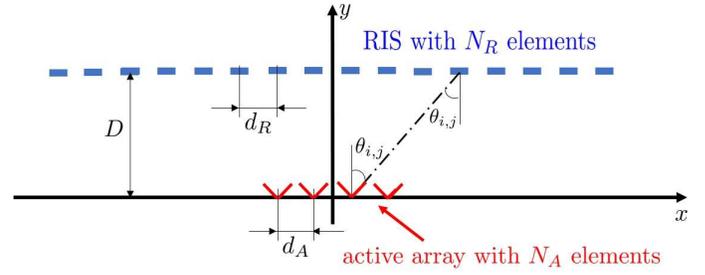}
	\end{center}
	\caption{Considered antenna scheme. $\theta_{i,j}$ is the look angle of the $i$-th element of the active array with respect to the $j$-th element of the RIS.}
	\label{fig:antenna}
\end{figure}

We model the RIS-to-active array $(N_A \times N_R)$-dimensional matrix $\bH$. To this aim, and for the sake of simplicity, we consider the 2D scheme represented in Fig. \ref{fig:antenna}, where we have an active array that illuminates a large RIS placed at a short distance, that we denote by $D$; the UEs to be served are placed in the backside of the active antenna array. We make the following assumptions:
\begin{itemize}
	\item[i)] We neglect the blockage that the active array can introduce on the electromagnetic radiation reflected by the RIS. In practical 3D scenarios, the active array can be placed laterally with respect to the RIS so as to simply avoid the blockage problem. 
	\item[ii)] The direct signal from the active antenna array to the mobile UEs will be neglected because it is considerably weaker than that from the RIS-reflected signal.
	\item[iii)] Given the short distance between the RIS and the active antenna array, the channel $\bH$ is modeled as a deterministic quantity.
	\item[iv)] We neglect the mutual coupling between the elements of the active array and between the elements of the RIS. This is an assumption that is usually verified provided that the element spacing at the active array and at the RIS, i.e. $d_A$ and $d_R$, respectively, does not fall below the typical $\lambda/2$ value \cite{qian2021mutual}, with $\lambda$ being the wavelength corresponding to the radiated frequency.  
	\item[v)] We assume that each element of the RIS is in the far field of each element of the active antenna array. However,  the whole RIS is not required to be in the far field of the whole active antenna array. 
\end{itemize}
With regard to assumption v) we note that the far field condition among the generic $i$-th transmit antenna and $j$-th RIS element is $d_{i,j}>2 \max\{ \Delta_A^2, \Delta_R^2\}/\lambda$ and 
$d_{i,j}\gg \max\{ \Delta_A, \Delta_R, \lambda\}$, with $\Delta_A$ and $\Delta_R$ the size of the active antenna and RIS element, respectively \cite{stutzman2012antenna}. The far-field condition among the whole RIS and the whole active array would have instead required that the fulfillement of the previous conditions with $d_{i,j}$,  $\Delta_A$ and $\Delta_R$ replaced by $D$, $N_A d_A$ and $N_R d_R$. The absence of the far field assumption between the whole active array and the RIS, although making the matrix $\bH$ modelling somewhat more involved, is crucial in order to have a full-rank matrix channel that will permit communicating along many signal space directions.  

Based on the above assumptions, and according to standard electromagnetic radiation theory, the $(i,j)$-th entry of $\bH$, say $\bH(i,j)$, representing the uplink channel coefficient from the $j$-th element of the RIS to the $i$-th active antenna, can be written as:
\begin{equation}
	\bH(i,j)=\ds \sqrt{\rho G_A(\theta_{i,j})G_R(\theta_{i,j})} \frac{\lambda}{4 \pi d_{i,j}} e^{-\i 2\pi d_{i,j}/\lambda} \; ,
	\label{eq:H_ij}
\end{equation}
where $G_A(\theta_{i,j})$ and $G_R(\theta_{i,j})$ represent the $i$-th active antenna and the $j$-th passive RIS element gains corresponding to the look angle $\theta_{i,j}$ (see Fig. \ref{fig:antenna}), while $\rho\leq1$ is a real-valued term modeling the RIS efficiency in reflecting the impinging waves. 

\subsection{Active antenna configuration}

In order to see the impact of the active antenna configuration, we consider two kinds of active antennas: \textit{omnidirectional} and \textit{directional}. In the first case, the active antennas are omnidirectional and radiate power in all the angular directions, i.e., $G_A(\theta)=3$ dB $\forall \theta \in [0, \pi]$. In the latter case, the active antennas radiate power only in one angular sector, i.e., $G_A(\theta)=g_{\text{dir}}$, $\forall \theta \in [-\alpha, \alpha]$. Note that, given the RIS and active array relative position depicted in Fig. \ref{fig:antenna}, if $N_R \gg N_A$ some RIS elements are external to the angular sector in which each active antenna radiate power. This situation strongly degrades the performance of the system because, actually, the ``effective'' elements of the RIS are only the ones that receive power from active antennas. To avoid this undesired behaviour, we can increase the active antenna spacing $d_A$ according to

\begin{equation}
	d_A=\ds \frac{(N_R-1)d_R-2D\tan{\alpha}}{N_A - 2} \,
	\label{spacing_directional}
\end{equation} 
where the above relationship can be obtained with simple geometric formulations and the notation refers to Fig. \ref{fig:antenna}.

\section{Uplink/Downlink signal processing}

\subsection{Uplink channel estimation}
During the uplink training phase, each UE transmits a known pilot sequence; we denote by $\bvphi_k$ the pilot sequence, of length $\tau_p$ and squared norm $\tau_p$, assigned to the $k$-th UE. Since, in this paper we assume $N_R>N_A$, in order to allow the active array to estimate the channels $\bh_1, \bh_K$, we use a similar approach as in \cite{Buzzi_TCCN_2021}. In particular, we assume that the pilot sequences are transmitted by the UEs $Q$ times, each one with a different RIS configuration. Without loss of generality, we assume $Q=N_R/N_A$ in order to guarantee generation of a number of observables that is not smaller than the number of unknown coefficients.  Pilots are drawn from a set of orthogonal sequences, which, for the sake of simplicity, are assumed to be real. 
Denoting by $\eta_k^{(u,t)}$ the uplink transmit power from the $k$-th UE during the training phase, the data observed at the active antenna array when the RIS assumes the $q$-th configuration can be arranged in the following $(N_A \times \tau_p)$-dimensional matrix:
\begin{equation}
	\bY^{(q)}=\ds \sum_{k=1}^K \sqrt{\eta_k^{(u,t)}}\bH \bPhi_T^{(q)} \bh_k \bvphi_k^T + \bN^{(q)} \; ,
\end{equation}
where $\bPhi_T^{(q)}$ is the $q$-th configuration of the RIS assumed
$
	\bPhi_T^{(q)}=\text{diag}\left(\bphi^{(q)}\right),
$
and $\bphi^{(q)}$ is a $N_R$-dimensional vector whose $\ell$-th entry is $e^{\i\widetilde{\phi}_{\ell,q}}$ with $\widetilde{\phi}_{\ell,q}$ a random angle chosen from a quantized set of phase shifts available at the RIS. 
In order to estimate the $k$-th UE channel $\bh_k$, the data $\bY^{(q)}$ is projected along the $k$-th UE pilot sequence, leading to 
\begin{equation}
	\by_{k,q}=\sqrt{\eta_k^{(u,t)}}\tau_p \bH\bPhi_T^{(q)} \bh_k + \sum_{\substack{j=1\\ j \neq k}}^K
	\sqrt{\eta_j^{(u,t)}} \bH \bPhi_T^{(q)} \bh_j \rho_{j,k} + \bn_{k,q} \; ,
	\label{eq:y_k}
\end{equation}	
with $\rho_{j,k}=\bvphi_j^T \bvphi_k$, and $\bn_{k,q} = \bN^{(q)} \bvphi_k \sim {\cal CN}(0,\sigma^2 \tau_p \bI_{N_A})$. These observables are collected for all $q=1,\ldots,Q$ as the following $N_R \times 1$-dimensional vector
$
	\widetilde{\by}_k=\left[ \by_{k,1}^T, \; \ldots , \; \by_{k,Q}^T\right]^T.
$
After some manipulations, $\widetilde{\by}_k$ can be written as
\begin{equation}
	\widetilde{\by}_k=\sqrt{\eta_k^{(u,t)}}\tau_p \widetilde{\bH}_T \bh_k + \sum_{\substack{j=1\\ j \neq k}}^K
	\sqrt{\eta_j^{(u,t)}} \widetilde{\bH}_T \bh_j \rho_{j,k} + \widetilde{\bn}_k \; ,
	\label{y_k_tilde2}
\end{equation}
where $\widetilde{\bn}_k=\left[ \bn_{k,1}^T, \; \ldots , \; \bn_{k,Q}^T\right]^T$ and
\begin{equation}
	\widetilde{\bH}_T=\left[ \left(\bH\bPhi_T^{(1)}\right)^T, \; \ldots , \; \left(\bH\bPhi_T^{(Q)}\right)^T\right]^T.
	\label{H_tilde_T}
\end{equation}

In order to estimate the \textit{essential directions} of the channels $\left\lbrace \bh_1, \ldots, \bh_K  \right\rbrace$, we use the economy size singular valued decomposition (SVD) of the matrix $\widetilde{\bH}_T$
\begin{equation}
	\widetilde{\bH}_T=\mathbf{U}\bm{\Lambda}\mathbf{V}^H \approx \widetilde{\mathbf{U}}\widetilde{\bm{\Lambda}}\widetilde{\mathbf{V}}^H,
	\label{H_tilde_T_svd}
\end{equation}
where $\mathbf{U},\bm{\Lambda},\mathbf{V}$ are the $N_R$-dimensional matrices corresponding to the regular SVD and, say $q$ the  number of significant eigenvalues of the matrix $\widetilde{\bH}_T$, $\widetilde{\mathbf{U}}$ and $\widetilde{\mathbf{V}}$ contains the columns of $\mathbf{U}$ and $\mathbf{V}$ corresponding to the $q$ highest eigenvalues and $\widetilde{\bm{\Lambda}}$ is the diagonal matrix containing these eigenvalues. Note that the approximation in Eq. \eqref{H_tilde_T_svd} holds if the matrix $\widetilde{\bH}_T$ has $q$ eigenvalues significantly higher with respect to the remaining ones. In this paper, it holds because of the structure of the matrix $\bH$. Given the properties of the SVD, we have
$
\bh_k= \mathbf{V} \mathbf{V}^H \bh_k = \mathbf{V} \mathbf{v}_k,
$
where $\mathbf{v}_k=\mathbf{V}^H \bh_k= \left[ \widetilde{\mathbf{v}}_k^T, \overline{\mathbf{v}}_k^T\right]^T$ and using the approximation in Eq. \eqref{H_tilde_T_svd} the following holds
\begin{equation}
	\bh_k \approx \widetilde{\mathbf{V}} \widetilde{\mathbf{v}}_k.
	\label{v_k_approx}
\end{equation}
We perform a channel estimation technique aimed at determining the strongest directions of the channels, corresponding to the highest eigenvalues, which we can estimate with highest power, and thereby recover the channels exploiting Eq. \eqref{v_k_approx}. Using the above considerations, we focus on the $q$-dimensional vector
\begin{equation}
	\begin{array}{lll}
		\overline{\by}_k= &\!\!\!\!\!\widetilde{\mathbf{U}}^H\widetilde{\by}_k\!\!\approx\!\!\sqrt{\eta_k^{(u,t)}}\tau_p \widetilde{\bm{\Lambda}}\widetilde{\mathbf{v}}_k\!\! +\!\!\!  \ds \sum_{\substack{j=1\\ j \neq k}}^K
		\!\!\sqrt{\eta_j^{(u,t)}} \widetilde{\bm{\Lambda}}\widetilde{\mathbf{v}}_j\rho_{j,k} \!+\! \widetilde{\mathbf{U}}^H\widetilde{\bn}_k \; .
	\end{array}
	\label{y_k_bar2}
\end{equation}
Thus, the linear MMSE estimate of $\widetilde{\mathbf{v}}_k$ can be expressed as
$
	\wh{\widetilde{\mathbf{v}}}_k=\bR_{\widetilde{\mathbf{v}}_k \overline{\by}_k}\bR_{\overline{\by}_k \overline{\by}_k}^{-1}\overline{\by}_k \; ,
$
with
\begin{equation}
	\bR_{\widetilde{\mathbf{v}}_k \overline{\by}_k} \triangleq E\left\{ \widetilde{\mathbf{v}}_k \overline{\by}_k^H \right\}=\sqrt{\eta_k^{(u,t)}} \tau_p \beta_k \widetilde{\bm{\Lambda}}^H, 
\end{equation}
\begin{equation}
	\bR_{\overline{\by}_k \overline{\by}_k}\!\! \triangleq \!E\left\{ \overline{\by}_k \overline{\by}_k^H \right\}\!\!=\!\!\Bigg[ \ds \sum_{j=1}^K \eta_j^{(u,t)} \beta_j  \rho_{j,k}^2\Bigg]\!  \widetilde{\bm{\Lambda}} \widetilde{\bm{\Lambda}}^H\!\! + \sigma^2 \tau_p \bI_{q}.	
\end{equation}
The channel estimate $\wh{\bh}_k$ can be thus written as 
\begin{equation}
	\wh{\bh}_k \approx \widetilde{\mathbf{V}} \bR_{\widetilde{\mathbf{v}}_k \overline{\by}_k}\bR_{\overline{\by}_k \overline{\by}_k}^{-1}\overline{\by}_k.
	\label{v_k_approx_estimate}
\end{equation}
It is easy to realize that the LMMSE estimation of the vector containing the strongest directions of the channel $\wh{\bh}_k$ is Gaussian distributed with zero mean and covariance matrix 
\begin{equation}
	\bR_{\wh{\bh}_k \wh{\bh}_k}=\widetilde{\mathbf{V}}\bR_{\widetilde{\mathbf{v}}_k \overline{\by}_k}\bR_{\overline{\by}_k \overline{\by}_k}^{-1}
\bR_{\widetilde{\mathbf{v}}_k \overline{\by}_k}^H \widetilde{\mathbf{V}}^H.
\end{equation}

\subsection{Downlink data transmission}
During the downlink data transmission phase, the complex envelope of the signal received by the $k$-th UE in the $n$-th symbol interval can be shown to be expressed as
\begin{equation}
	\begin{array}{lll}
		r_k(n)=&\sqrt{\eta_k^{(d)}} \bh_k^T \bPhi^T \bH^T \bw_k x_k(n)  \\ &\ds +\sum_{\substack{j=1\\ j\neq k}}^K
		\sqrt{\eta_j^{(d)}} \bh_k^T \bPhi^T \bH^T \bw_j x_j(n) + z_k(n) \, ,
	\end{array} \label{eq:downlink_data}
\end{equation}
where $\eta_k^{(d)}$ and $\bw_k$ represent the downlink transmit power and the downlink beamformer for UE $k$, respectively, while $x_k(n)$ is the data symbol, drawn from a constellation with unit average energy, intended for UE $k$ in the $n$-th signaling interval. The term $z_k(n)$ is the AWGN term, modeled as ${\cal CN}(0, \sigma^2_k)$. 

\section{Performance measures} \label{Performance_measures}
For benchmarking purposes, we start considering the ideal case in which perfect CSI is available both at the active array and at the UEs.
Given \eqref{eq:downlink_data}, the $k$-th UE downlink SINR, for perfect CSI, is written as
\begin{equation}
	\gamma_k^{(d)}= \ds \frac{\eta_k^{(d)} \left| \bh_k^T \bPhi^T \bH^T   \bw_k\right|^2}{
		\ds \sum_{\substack{j=1\\ j\neq k}}^K \eta_j^{(d)} 	\left| \bh_k^T \bPhi^T \bH^T \bw_j \right|^2 + \sigma^2_k  \; .
	}
	\label{eq:DL_SINR}
\end{equation}
Consequently, assuming Gaussian-distributed codewords, the spectral efficiency (SE) for the $k$-th UE  can be obtained by using the classical Shannon expression, i.e.:
\begin{equation}
	\text{SE}_k^{(d)}= \xi \log_2 (1+ \gamma_k^{(d)}) \; , \quad [\mbox{bit/s/Hz}] \; .
	\label{DL_SE}
\end{equation}
where $\xi$ accounts for the fraction of the channel coherence interval used for the downlink data transmission.
In general, when the channel realizations are not perfectly known, the UE cannot clearly tell what is ``signal'' and what is
``interference'' in Eq. \eqref{eq:downlink_data}. With imperfect CSI, the intuitive notion of SINR in Eq. \eqref{eq:DL_SINR} is in general not rigorously related to a corresponding notion of information theoretic achievable rate \cite{Caire_Bounds2018}. 

\subsection{The upper-bound (UB) to the system performance}
An upper bound of the downlink SE, can be obtained by optimistically assuming perfect knowledge of the channel estimates at the UEs. This bound gives us a first sense of the system performance, and is given by~\cite{Caire_Bounds2018}
\begin{equation}
	\text{SE}_{k,\rm{UB}}^{(d)}= \bar{\xi}~E\left\{\log_2 (1+ \gamma_k^{(d)})\right\} \; , \quad [\mbox{bit/s/Hz}] \; ,
	\label{rate_Dl_UB}
\end{equation}
where the pre-log factor $\bar{\xi}<1$ accounts for the fraction of the channel coherence interval used for downlink data transmission, and for the pilot overhead. In Eq. \eqref{rate_Dl_UB} the expectation is computed over the small-scale fading quantities, and the beamforming vectors $\bw_1,\ldots, \bw_K$ in Eq. \eqref{eq:DL_SINR} are designed upon the channel estimates.

\subsection{The hardening lower-bound (LB) to system performance} \label{LB_performance}
To obtain an achievable SE expression in closed form, we derive the so-called \textit{hardening} lower-bound on the capacity, as detailed in~\cite{redbook}.
Under the assumption of Rayleigh fading, and using a conjugate beamformer, i.e. letting
$\bw_k=\bH^* \bPhi^* \widehat{\bh}_k^*$, an achievable SE for the $k$-th UE can be obtained by treating all the interference and noise contributions as uncorrelated effective noise. Letting,
\begin{equation}
	\begin{array}{lll}
		&r_k(n)=\underbrace{\sqrt{\eta_k^{(d)}} E\left\{\bh_k^T \bPhi^T \bH^T \bw_k\right\}}_{\text{DS}_k} x_k(n) 
		\\ & + \ds \underbrace{\sqrt{\eta_k^{(d)}}\left[\bh_k^T \bPhi^T \bH^T \bw_k - E\left\{\bh_k^T \bPhi^T \bH^T \bw_k\right\}\right]}_{\text{BU}_k} x_k(n)
		\\ &\ds +\sum_{\substack{j=1\\ j\neq k}}^K
		\underbrace{\sqrt{\eta_j^{(d)}} \bh_k^T \bPhi^T \bH^T \bw_j}_{\text{UI}_{k,j}} x_j(n) + z_k(n) \, ,
	\end{array} 
	\label{eq:downlink_data_hardening}
\end{equation}
then the hardening lower-bound is given by
\begin{equation}
	\text{SE}_{k,\rm{LB}}^{(d)} =  \bar{\xi} \log_2(1+ \gamma_{k, \text{LB}}^{(d)}) \, ,
	\label{eq:hardening}
\end{equation} \vspace*{-2mm}
with 
\begin{equation}
	\gamma_{k, \text{LB}}^{(d)}= \ds\frac{\left|\text{DS}_k\right|^2}{
		E\{\left|\text{BU}_k\right|^2\} + \ds \sum_{\substack{j=1\\ j\neq k}}^K E
		\{\left| \text{UI}_{k,j} \right|^2 \}+ \sigma^2_k }.
\end{equation}

The closed-form expression of $\gamma_{k, \text{LB}}^{(d)}$ is reported in Eq. \eqref{eq:SINR_D_LB} at the top of next page, where $\mathcal{P}_k$ includes the indices of the UEs using the same pilot as user $k$.
Due to space constraints, we omit the proof including the derivation of~\eqref{eq:SINR_D_LB}.
\begin{figure*}[t]
	\begin{equation}
		\gamma_{k, \text{LB}}^{(d)}= \ds \frac{{\eta_k^{(d)}} \trace^2\left[   \APhi \bR_{\wh{\bh}_k \wh{\bh}_k}^*\right]}
		{\left\lbrace\begin{array}{lr}
				\eta_k^{(d)} \trace\left( \beta_k \APhi \bR_{\wh{\bh}_k \wh{\bh}_k}^* \APhi\right) + \ds \sum_{j \in {\cal P}_k} \eta_j^{(d)}\left(
				\ds \frac{\beta_j \sqrt{\eta_j^{(u,t)}}}{\beta_k \sqrt{\eta_k^{(u,t)}}}\right)^2 \left[ 
				\trace^2(\APhi \bR_{\wh{\bh}_k \wh{\bh}_k}^*)+ \beta_k\trace(\APhi \bR_{\wh{\bh}_k \wh{\bh}_k}^* \APhi)
				\right]    +  \\  
				\hfill \ds \sum_{j \notin {\cal P}_k}    \eta_j^{(d)} \beta_k\trace\left[ \APhi 
				\bR_{\wh{\bh}_j \wh{\bh}_j}^*\APhi\right]
				+ \sigma^2_k \end{array}\right \rbrace}  	
	\label{eq:SINR_D_LB}
	\end{equation}
	\hrulefill
\end{figure*}

Finally, we can note that the analysis of the considered structure is, from a mathematical point of view, different from that required to analyze conventional massive MIMO systems with one active antenna array with a large number of elements.
As it is well known, in a conventional massive MIMO system, when the number of antenna elements grows without bound we can observe two phenomena: the \textit{channel hardening} (the channel norm tends to be deterministic) and the \textit{favorable propagation} (channel vectors of two different UEs become almost orthogonal). 
In the case of RIS-aided massive MIMO, letting $N_R$ grow without bound and keeping $N_A$ fixed does not provide the above described effects. It can be shown that these properties hold \textit{only} if the number of active antennas increases without bound, we omit this proof here due to the lack of space. 

\section{Optimization of the RIS phase shifts} \label{RIS_optimization}
In order to optimize the system performance for finite $N_A$, we define the matrix
$
\mathbf{Q}\left( \bPhi\right)=\mathbf{S}\left( \bPhi\right)\mathbf{S}\left( \bPhi\right)^H,
$
with
$
\mathbf{S}\left( \bPhi\right)=\bH \bPhi \overline{\bH},
$
and $\overline{\bH}=\left[ \bh_1, \ldots, \bh_K\right ]$.
The entries of matrix $\mathbf{Q}\left( \bPhi\right)$ give us an idea on the potential interference that the UEs suffer for a given RIS configuration $\bPhi$.
We thus focus on the matrix $\mathbf{Q}\left( \bPhi\right)$ in order to formulate two problems aimed at the optimal configuration of the RIS phase offsets. We define two objective functions:
\begin{equation*}
f_1\left( \bPhi\right)= \ds \sum_{k=1}^K \sum_{k'=k+1}^K \!\!\! |\bar{\bh}_k^H \bar{\bh}_{k'}|, \; \text{and} \; f_2\left( \bPhi\right)= \ds \frac{f_1\left( \bPhi\right)}{\ds \sum\nolimits_{k=1}^K \|\bar{\bh}_k\|^2} \, ,
\end{equation*}
where 
$
\bar{\bh}_k=\bH \bPhi \bh_k \, 
$
and the functional dependence on $\bPhi$ is omitted for ease of notation.
The minimization problem with objective function $f_1\left( \bPhi\right)$ is aimed at the minimization of the cross-correlation among channel vectors to reduce interference. The one with objective function $f_2\left( \bPhi\right)$ strikes a balance between the conflicting requirements that the RIS phases should not lower the norms of the composite channels, while at the same time should minimize cross-correlation among channel vectors. 
The generic optimization problem can be written as
\begin{subequations}\label{Prob:RIS_opt}
\begin{align}
&\ds\min_{\bPhi}\;  f_{(\cdot)}\left( \bPhi\right)\\
&\;\textrm{s.t.}\; |\phi_i|=1,\; i=1,\ldots, N_R
\end{align}
\end{subequations}
where $(\cdot) \, \in \left\lbrace 1,2 \right\rbrace$. 
The optimization problem \eqref{Prob:RIS_opt} is not convex, hence we propose a solution based on the alternating maximization theory\cite{BertsekasNonLinear}.
We consider the $N_R$ optimization problems, $i=1,\ldots, N_R$,
\begin{subequations}\label{Prob:RIS_opt_ith}
	\begin{align}
		&\ds\min_{\phi_i}\;  f_{(\cdot)}\left( \phi_i\right)\\
		&\;\textrm{s.t.}\; |\phi_i|=1,
	\end{align}
\end{subequations}
with $\phi_{\ell}, \ell=1,\ldots, i-1,i+1,\ldots, N_R$ remain fixed. This optimization problem can be solved using an exhaustive search. It is proved in \cite[Proposition 2.7.1]{BertsekasNonLinear} that iteratively solving \eqref{Prob:RIS_opt_ith} monotonically decreases the value of the objective of \eqref{Prob:RIS_opt}, and converges to a first-order optimal point if the solution of \eqref{Prob:RIS_opt_ith} is unique for any $i$. The generic optimization algorithm can be summarized as in Algorithm \ref{Alg:RIS_opt}.
\begin{algorithm}[t]
	\caption{Optimization of the RIS phase shifts}
	\begin{algorithmic}[1]
		\label{Alg:RIS_opt}
		\STATE Choose a starting random RIS configuration of the phase shifts $\bPhi$;
		\REPEAT
		\FOR{$i=1\to N_R$}
		\STATE Solve Problem \eqref{Prob:RIS_opt_ith};
		\STATE Set the $(i,i)$-th entry of the matrix $\bPhi$ as the solution of Problem \eqref{Prob:RIS_opt_ith}, $\phi_i^*$ say.
		\ENDFOR
		\UNTIL{convergence of the objective function}
	\end{algorithmic}
\end{algorithm}

\begin{figure*}[t]
	\begin{center}	\includegraphics[scale=0.41]{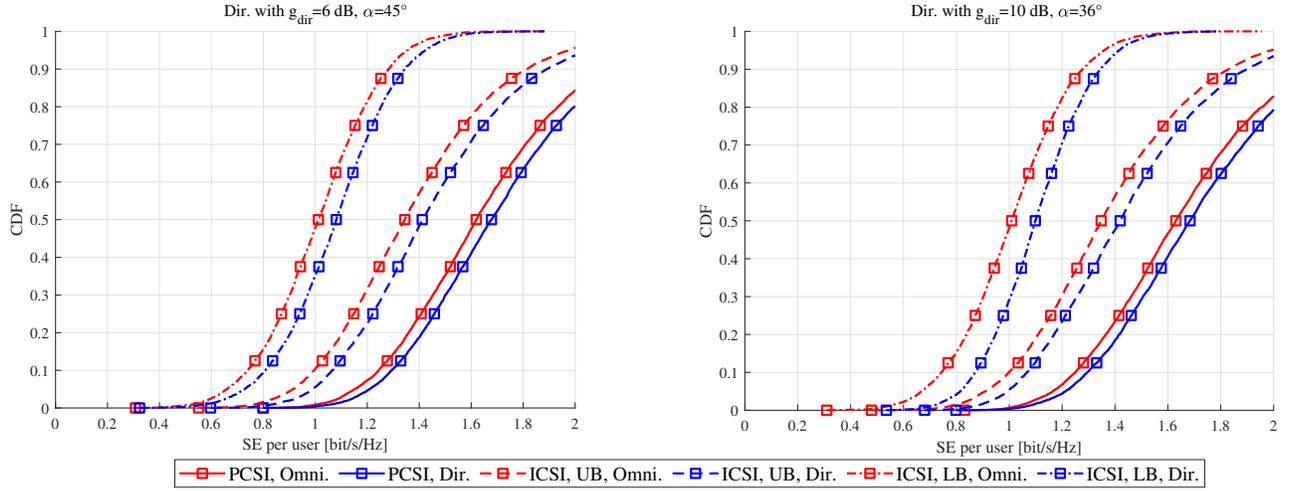}
	\end{center}
	\caption{CDF of the Spectral efficiency (SE) per user in the case of omnidirectional (Omni.) and directional (Dir.) active antennas with $N_R=64$ and $N_A=16$. Comparison between the case of perfect CSI (PCSI), and imperfect CSI (ICSI) with upper-bound (UB) and lower-bound (LB) of the performance.}
	\label{fig:RIS_bounds}
\end{figure*}

\section{Numerical Results}
In our simulation setup, we consider a communication bandwidth of $W = 20$ MHz centered over the carrier frequency
$f_0 = 1.9$ GHz. The additive thermal noise is assumed to have a power spectral density of $-174$ dBm/Hz,
while the front-end receiver at the active array and at the UEs is assumed to have a noise figure of 5 dB. The heights of active array and RIS are 10 meters while the heigh of the UEs is 1.5 meters. The $K=8$ UEs are uniformly distributed in an angular region of $120 \deg$ in front of the RIS at distances in $[10,400]$ meters. We model the large scale fading coefficients $\beta_k$ from the UEs to the RIS according to \cite{3GPP_38901_model} and the matrix $\bH$ according to Eq. \eqref{eq:H_ij} with $D=5\lambda$, $\rho=1$, $d_A=\lambda/2$ and in the case of omnidirectional active antennas $G_R\left( \theta \right) = 3$ dB $\forall \theta$, $d_R=\lambda/2$; in the case of directional active antennas  $d_R$ follows Eq. \eqref{spacing_directional} with $g_{\text{dir}}$ and $\alpha$, specified in the figures.  With regard to the channel estimation procedure, $q$ is chosen as the number of eigenvalues which collects the 98\% of the total sum of eigenvalues. The pilot sequences have length $\tau_p = 16$, and the uplink transmit power for the pilot is assumed to be $\eta_k=800$ mW. The coherence interval is $\tau_c = 200$ samples long, and $\tau_c-\tau_p$ samples are used for downlink data. This implies $\bar{\xi}=1-\tau_p/\tau_c$. The maximum transmit power of the active array is $P_{\rm max}^{{\rm A}}=7$ dBW and it is equally distributed between the UEs, i.e., $\eta_k^{(d)}= P_{\rm max}^{{\rm A}}/K \norm{\bw_k}^2$.

\begin{figure*}[t]
	\begin{center}	\includegraphics[scale=0.41]{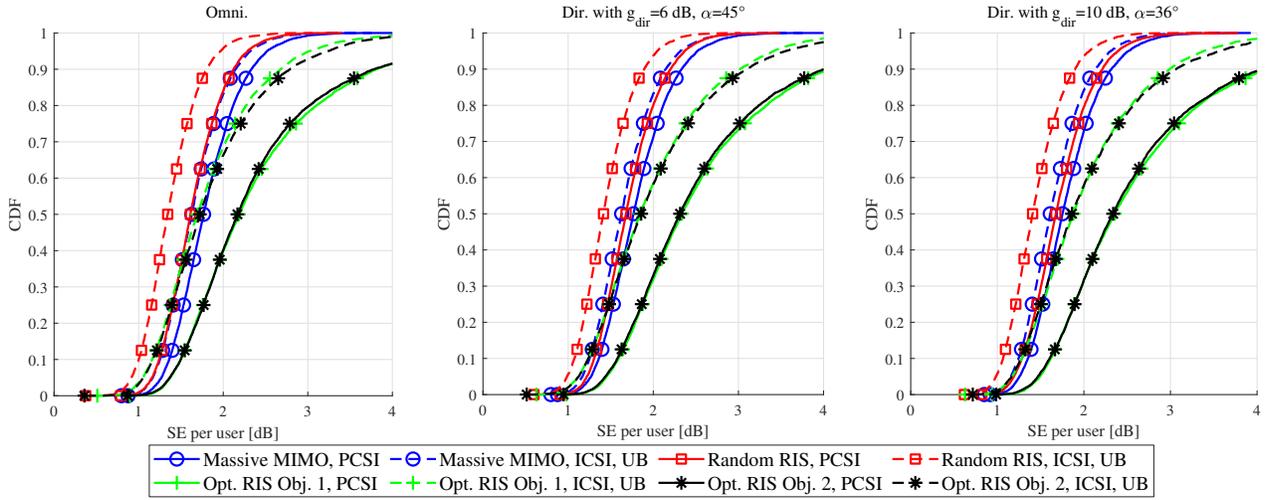}
	\end{center}
	\caption{CDF of the Spectral efficiency (SE) per user in the case of omnidirectional (Omni.) and directional (Dir.) active antennas with $N_R=64$ and $N_A=16$. Comparison between the case of perfect CSI (PCSI), and imperfect CSI (ICSI) with upper-bound (UB) of the performance with RIS phase shifts optimization with objective functions $f_1(\cdot)$ and $f_2(\cdot)$.}
	\label{fig:RIS_optimization}
\end{figure*}

In Fig. \ref{fig:RIS_bounds} we report the cumulative distribution function (CDF) of the spectral efficiency (SE) per user using the expressions discussed in Section \ref{Performance_measures} for perfect CSI (PCSI) and imperfect CSI (ICSI). In this figure, we assume a random configuration of the RIS phase shifts. We can see that in the case of directional active antennas the performance in terms of SE increases with respect to the omnidirectional case due to the higher values in the matrix $\bH$ which models the communication between the RIS and the active array. With the considered system, we can note that the hardening LB is not as tight as in the case of classical massive MIMO because of the reasons discussed in Section \ref{LB_performance}.
In Fig. \ref{fig:RIS_optimization} we report the CDF of the SE per user in the case of PCSI and ICSI using the UB of the performance. We report the cases of RIS with random configuration of phase shifts (Random RIS) and in the cases of phase shift optimization (Opt. RIS) detailed in Section \ref{RIS_optimization} where we denote by Obj. 1 the Problem \eqref{Prob:RIS_opt} with objective function $f_1(\cdot)$ and Obj. 2 the one with objective function $f_2(\cdot)$. We compare the performance with the traditional Massive MIMO with $N_A$ antennas. We can see that the optimized RIS with directional active antennas strongly increases the performance of the system compared with the traditional massive MIMO, especially in the PCSI case. Additionally, we can also see that the proposed channel estimation procedure is effective and that the phase shift optimization works well also in this case, especially when directional active antennas are assumed.  

\section{Conclusions}
We considered a RIS-aided massive MIMO system in which a RIS is placed at short distance and in front of an active antenna array, with a non-large number of elements. The UEs to be served are placed in the backside of the active antenna array. We provided a signal and channel model suitable for the analysis of such a system and proposed a channel estimation procedure which, using different configurations of the RIS phase shifts, is capable to estimate the channel from the large number of RIS elements and all the UEs exploiting only the active antennas. We derived a closed-form expression for for an achievable downlink SE per user, by using the popular hardening lower-bound, and formulated a generic optimization problem which can be used for two purposes. In the numerical results, we showed the effectiveness of the considered architecture assuming both omnidirectional and directional active antennas, and of the phase shifts optimal configuration algorithm.

\bibliographystyle{IEEEtran}
\bibliography{biblio}

\end{document}